\documentclass[11pt]{article}  
\usepackage{graphicx}
\usepackage{amssymb}
\usepackage{color}

\input epsf
\parskip=\medskipamount
\def\la{\lambda}  
\def\kp{\kappa}  
\def\te{\theta}   
   
\def\IB{\relax{\rm l\kern-.18 em B}}
\def\IC{\relax{\rm l\kern-.50 em C}}
\def\IK{\relax{\rm l\kern-.10 em K}}
\def\IL{\relax{\rm I\kern-.18 em L}}
\def\IN{\relax{\rm I\kern-.18 em N}}
\def\IR{\relax{\rm I\kern-.18 em R}}
\def\ii{\rm i\,}
\def\smallonehalf{\frac{{}_1}{{}^2}}
 
\def\\{\hfill\break}
\def\smallonehalf{\frac{{}_1}{{}^2}}
 
\def\tr{\mathop{\rm tr}\nolimits}

\font\tenfrak=eufm10  \font\sevenfrak=eufm7  \font\fivefrak=eufm5
\newfam\frakfam
\textfont\frakfam=\tenfrak
\scriptfont\frakfam=\sevenfrak
\scriptscriptfont\frakfam=\fivefrak

\newtheorem{proposition}{Proposition}

\def\frac#1#2{{#1\over #2}}

\def\ptos{\leaders\hbox to 2mm{\hfil{.}\hfil}\hfill}

\begin{document}

\title{ Superintegrability of 3-dimensional   Hamiltonian systems 
with conformally Euclidean metrics.  Oscillator-related and Kepler-related systems  
}

\author{ 
Jos\'e F.\ Cari\~nena$\dagger\,^{a)}$,
Manuel F.\ Ra\~nada$\dagger\,^{b)}$, and
Mariano Santander$\ddagger\,^{c)}$ \\ [2pt]
$\dagger$
   {\sl Departamento de F\'{\i}sica Te\'orica and IUMA, Facultad de Ciencias} \\
   {\sl Universidad de Zaragoza, 50009 Zaragoza, Spain}  \\   [2pt]
$\ddagger$
   {\sl Departamento de F\'{\i}sica Te\'{o}rica and IMUVa, Facultad de Ciencias} \\
   {\sl Universidad de Valladolid, 47011 Valladolid, Spain} 
} 
\date{}
\maketitle 

\begin{abstract}
We study  four particular 3-dimensional natural   Hamiltonian systems defined in  conformally Euclidean  spaces.
We prove  their superintegrability and we obtain, in the four cases,  the  maximal number of functionally independent integrals of motion. 
The two first systems are related  to the 3-dimensional isotropic oscillator and the superintegrability is quadratic. 
The third system is obtained as a continuous deformation  of an oscillator with ratio of frequencies 1:1:2 and with three additional nonlinear terms of the form  $k_2/x^2$,  $k_3/y^2$  and $k_4/z^2$, and the fourth system is obtained as a deformation of the Kepler Hamiltonian also with these three particular nonlinear terms.
These third and fourth systems are superintegrable but with higher-order constants of motion.

The four systems depend  on a real parameter in such a way that they are continuous functions of the parameter (in a certain domain of the parameter) and in the limit of such parameter going to zero the Euclidean dynamics is recovered.

\end{abstract}

\begin{quote}

{\sl Running title:}{\enskip}
Superintegrability of 3-dimensional   Hamiltonian systems with conformally Euclidean metrics.

\end{quote}

\vfill
\footnoterule{\small
\begin{quote}
$^{a)}${\it E-mail address:} {jfc@unizar.es } \\
$^{b)}${\it E-mail address:} {mfran@unizar.es }  \\
$^{c)}${\it E-mail address:} {mariano.santander@uva.es }
\end{quote}
}

\newpage

 \tableofcontents

\section{Introduction}

The problem of integrability of a given system of differential equations is a very interesting and active field of research along the last years  (see e.g. \cite{CFGR}). The 
most satisfactory situation would be the integrability by quadratures. 
By such  integrability we mean the possibility of  finding  the general solution in an algorithmic way,  and in this task the existence of additional structures, for instance compatible symplectic structures, may be useful. In the geometric approach   an autonomous system of first order differential equations is replaced by a vector field $X$ on a manifold $M$ in such a way that  the system is used to compute in a local coordinate system the integral curves of  the vector field. 
The determination of the integral curves of  $X$ is not an easy task and  the knowledge of its infinitesimal symmetries is very helpful. With this aim  one uses to look for invariant under $X$  geometric structures.  
For instance the complete  Liouville  integrability is developed in the framework of  symplectic, and more generally Poisson, structures.  
So, if we consider a $2n$-dimensional symplectic manifold $M$, for instance the 
cotangent bundle $T^*Q$ endowed with its natural symplectic structure $\omega_0$, completely integrable systems are defined by vector fields $X\in \mathfrak{X}(M)$ admitting a set of $n$ first integrals for $X$ in involution, giving rise to a Lagrangian foliation of $M$ and then using appropriate action-angle variables  we can carry out the determination of the integral curves of $X$.   

If there are more than $n$ functionally independent first integrals  for the vector field $X$ we say that the system is superintegrable  and moreover when  there  exists the maximum number, i.e.  $2n-1$, of  functionally independent first integrals we say that $X$ is  maximally superintegrable.   
There are not so many maximally superintegrable systems known,  but the importance of identifying such systems is  strengthened not only by their interesting mathematical properties but also by the fact that they can be used as approximation to non-integrable systems. 

Let us now consider two  conformally related vector fields $X$ and $f\, X$, where $f$ is a non vanishing function. 
They have the same constants of motion and therefore the integral curves of $f\, X$ are obtained from those of $X$ by a different reparametrization of each orbit. 
In fact this change of the vector field $X$ by a conformal one $f\, X$ corresponds to a generalization of the so-called  Sundman transformation  \cite{Sud13}, or infinitesimal time reparametrization,   which had previously been used by Levi-Civita \cite{LC04,LC06}, in the theory of  differential  equations
 \begin{equation}   
    dt= r\, d\tau\,,  \label{Sundman}  
\end{equation} 
 but now in this generalization the radial coordinate  $r$ is replaced by an  arbitrary function $f$ of the position coordinates. 
     
    We are interested in the case of a Lagrangian system.  First we note that the new velocities $\bar v^i$  must be  obtained by making use of the new time $\tau $ and therefore they do not coincide with  $v^i$, but are given by  $\bar v^i=f\,  v^i$. 
Moreover, in addition we recall that  in the framework of the Lagrangian formalism  the relevant concept is the action defined by the Lagrangian,  and in order to preserve the action,  if a system was defined by a Lagrangian $L$, then the new system must be described in terms of the new time $\tau$ by a new Lagrangian $\overline L(q,\bar v)$ given by 
$$  \overline L(q,\bar v) = f\, L\left(q,\frac{\bar v} f\right).
$$
In the particular case of a free motion on a Riemann manifold $(Q,g)$, where the Lagrangian $L$ is just  the $g$-dependent kinetic energy function $T_g\in C^\infty(TQ)$ given by 
\begin{equation}
 T_g(v)=\frac{1}{2}\,g(v,v),\qquad v\in TM, \label{defTg}
\end{equation} 
then   the new Lagrangian will be the  kinetic energy determined by the metric $(1/f)\,g$ because $g$ is quadratic in velocities and we have therefore $f\, L(q,\bar v/f)=(1/f) L(q,\bar v)$.
For a system of   mechanical type, described by a Lagrangians $L=T_g-\tau^*V$ (where $V$ is a potential function defined  on the configuration space $Q$), the considered generalized Sundman transformation amounts to change not only the Riemann structure from $g$ to $\bar g=(1/f) g$ but also the potential function $V$ to $\overline V=f\, V$, and when passing to the Hamiltonian formalism, by making use of the Legendre transformation, the Hamiltonian $H$ of the  mechanical type system must be replaced by $\overline H= f\, H$. 

The usefulness of such a correspondence has been shown in many examples. For instance, a mechanical type system for which there is a coordinate system such that  the potential function is a sum 
$V(q)=V_1(q_1)+\ldots+V_n(q_n)$  and the local expression of the Riemann structure is diagonal, i.e.
$$ 
 L(q,v)=\frac 12 \sum _{i=1}^n a_i(q_i) v_i^2-  \sum _{i=1}^n V_i(q_i),
$$
is separable as a sum of one-dimensional systems  and hence integrable by quadratures. 
A generalisation of such system is due to Liouville  \cite{Liouv49} and consists on the Hamiltonian 
\begin{equation}
 H(q,v)=\frac 1{2W(q)} \sum _{i=1}^n a_i(q_i) v_i^2+  \frac 1{W(q)}\sum _{i=1}^n V_i(q_i),
\end{equation}
where $W(q)=W_1(q_1)+\ldots+W_n(q_n)$. 
These systems are called Liouville systems \cite{Perelomov,GzLMateos19} and one can check that the $n$ functions 
$$ 
 F_i=\frac 12 a_i(q)\, p_i^2+V_i(q)-W_i\,H,  \quad i=1,\ldots,n,
$$
are constants of motion $\{H, F_i\}=0$, but they are not independent because $\sum_{i=1}^n F_i=0$.

 Therefore, starting from an appropriate Hamiltonian $H$ we will analyze,  inspired by these results,  the possible functions  $f$ such that the new Hamiltonian $f H$ satisfies the required properties. 
 Note, however,  that $X_{f\, H}$ is different from $f X_ H $, that is  $X_{f\, H}=f X_H + Y$, the difference being the vector field $Y$ such that $i(Y)\omega_0=H\, df$. 
 We also note that, in the general case, $f X_ H $ is not a Hamiltonian vector field.

It is known that systems that admit Hamilton-Jacobi (Schr\"odinger in the quantum case) separability in more than one coordinate system are superintegrable with quadratic in the momenta constants of motion. 
In fact the modern studies on superintegrability started with ref. \cite{FrMS65} (probably the oldest study on this matter was the theorem of Bertrand \cite{Bertrand} although of course without using this word) in which the authors  proved the existence in the Euclidean plane of four families of potentials separable into two different sets of coordinates; two of them were related with the harmonic oscillator and the other two with the Kepler problem; 
in fact most of the superintegrable known systems (but not all) are related with these two important systems. 
Later on different authors have considered this question from different points of view. 
The 3-dimensional Euclidean systems with multiple separability and quadratic integrals were first studied by Evans \cite{Ev90PR} and then other different systems  were studied in different situations as on  two-dimensional pseudo-Euclidean spaces \cite{Ra97Jmp,Camp14Jmp}, on spaces with constant curvature \cite{GrPS95b}--\cite{GonGon20},  and even on more general curved spaces \cite{KaKW02}--\cite{CHR17Jmp}   (see \cite{MillPW13} for a review). 
We also note that the multiple separability of some Hamiltonian systems with linear terms in the momenta has also been studied  \cite{MarSno20,BeKuSn21}.  

Until recently, most studies on superintegrability were concentrated on the quadratic case but in these last years  the existence of systems possessing  integrals of motion of higher-order in the momenta (not arising from separability) has also been studied \cite{PPW12Jmp}--\cite{MarqWint19} but mainly in the two-dimensional Euclidean space. 

This paper is devoted to the study of  some superintegrable systems on 3-dimensional conformally Euclidean spaces (see  \cite{KKM05Jmp}--\cite{Fordy20}  for papers on this particular geometry). 
It is mainly concerned with systems related to the  harmonic oscillator and the Kepler problem. 

Suppose we are given a Hamiltonian $H$ of mechanical  type (quadratic kinetic term plus a potential function); then we can construct a new Hamiltonian  $ H_\mu$ as $H_\mu = \mu\, H$ where $\mu$ is a certain function defined on the configuration space. This new Hamiltonian represents a new and different  dynamics; for example if $H$ is defined on an Euclidean space then the new dynamics will be conformally Euclidean. 
The important point is that we are interested in multipliers $\mu$ that preserve certain properties as Liouville integrability. 
 A strong requirement is that $\mu$ must modify the dynamics but preserving  not just integrability but superintegrability. 
 For example, if $H$  is separable in Cartesian coordinates and $\mu$ is of the form $\mu=1/f$, $f = f_1(x) + f_2(y)+f_3(z)$, then $H_\mu$ is separable in Cartesian coordinates as well, and if $H$ is separable in spherical coordinates and $\mu$ is of the form $\mu=1/f$, $f = f_1(r) + f_2(\te)/r^2 + f_3(\phi)/(r^2\sin^2\te)$, then $H_\mu$ is also separable in $(r,\te,\phi)$ coordinates. 
 A more strong condition is that $\mu$ must preserve not just separability but multiple separability; this requirement will strongly restrict the form of the multiplier {(see \cite{Ran15Jmp} for a similar problem in the two dimensional case). 
 
 An important property to be imposed is that the new Hamiltonian $H_\mu$ must be a deformation of the original Hamiltonian $H$. By deformation we mean that $\mu$, and therefore $H_\mu$, will depend of a parameter $\la$ in such a way that 
\begin{itemize}
\item[(i)] The new Hamiltonian $H_\mu$  is a continuous function of $\la$ (in a certain domain of the parameter). 
\item[(ii)]  When $\la\to 0$ we have $\mu\to 1$ and then the dynamics of the Euclidean Hamiltonian $H$ is recovered.
\end{itemize}

Next we summarize the contents of this paper.

We study four different Hamiltonian systems with conformally Euclidean metrics and depending  on a continuous way of a real parameter $\la$.

In the four cases the potential of the original Euclidean system is a linear combination of four functions; the potential of the oscillator or the potential of the Kepler problem as the first and dominant term modified by the presence of three additional functions. 
First, in  Sec. (\ref{Sec2}) we study two different systems related with the 3-dimensional isotropic oscillators; we prove  their superintegrability and we obtain the explicit expression of six $\la$-dependent quadratic constants of motion (five of them functionally independent). 
Second, in Sec.  (\ref{Sec3})  we study a Hamiltonian obtained as a continuous deformation  of an oscillator with ratio of frequencies 1:1:2 and with three additional nonlinear terms of the form  $k_2/x^2$,  $k_3/y^2$  and $k_4/z^2$. 
We prove that this system is superintegrable with integrals of motion of fourth-order in the momenta.  
Third, in Sec. (\ref{Sec4}) we analyze a Kepler system modified with the three nonlinear terms $k_2/x^2$,  $k_3/y^2$  and $k_4/z^2$.  
Also in this system we obtain quartic constants of motions. 
In the four cases, the  multipliers $\mu$, leading to the conformally Euclidean systems, are  directly related with the first term of the potential, that is, harmonic oscillator in  Sec. (\ref{Sec2}) and (\ref{Sec3}) and Kepler potential in Sec. (\ref{Sec4}).

We obtain, in all the cases,  the value of the  sectional and Ricci curvatures of the metrics.

Finally we present an Appendix with the properties of a $\la$-dependent version of the Fradkin tensor constructed with the integrals of motion of one of the oscillators studied in Sec. (\ref{Sec2}). 

\section{Harmonic oscillator related Hamiltonians with a conformally Euclidean metric}  \label{Sec2}

\subsection{Isotropic oscillator with additional linear terms $k_2x$, $k_3y$, and $k_4z$}
\label{Sec21}

Let us consider the Hamiltonian  $H_{111}$ of the three-dimensional isotropic oscillator with additional terms of the form $k_2x$, $k_3y$, and $k_4z$
\begin{equation}
   H_{111} = ({\smallonehalf})\,\bigl( p_x^2 + p_y^2 +p_z^2 \bigr)  
 +  \bigl[\, k_1(x^2+y^2+z^2) + k_2 x +  k_3 y + k_4 z  \,\bigr]     \,,   
\end{equation}
and denote by  $\mu$ the following multiplier
\begin{equation}
 \mu  = 1/(1 -\la\,r^2) \,,{\quad} r^2= x^2+y^2+z^2 \,, 
\end{equation}
where $\la$ is a real parameter that can take both positive and negative values.
Then, the new $\la$-dependent Hamiltonian $H_\mu$ defined as 
$$
 H_{\mu} = \mu H_{111} = \bigl(\frac{1}{1-\la\, r^2}\bigr) H_{111} \,,{\quad} 
 \lim{}_{\la\to 0}  H_{\mu} =  H_{111}  \,, 
$$
takes the form 
\begin{equation}
  H_\mu = \mu H_{111}  =  ({\smallonehalf})\,\Bigl( \frac{p_x^2 + p_y^2 +p_z^2}{1 - \la\,r^2} \Bigr)  
 +  \Bigl[\,  k_1\frac{x^2+y^2+z^2}{1 - \la\,r^2} + \frac{k_2\, x }{1 - \la\,r^2} +  \frac{k_3\, y}{1 - \la\,r^2} + \frac{k_4\, z }{1 - \la\,r^2}  \,\Bigr]      \,. 
\end{equation}
 In the $\la<0$ case the dynamics of $H_\mu$  is correctly defined for all the values of the variables; 
nevertheless when $\la>0$, the Hamiltonian (and the associated  dynamics)
has a singularity at $1 -\,\la\,r^2=0$, so in this case the  $H_\mu$ dynamics is defined in the interior of the circle $r^2=1/\la$, $\la>0$,   which is the region where the  kinetic term is positive definite.

Notice that the complete potential in the Hamiltonian $H_{111}$ can be seen as an isotropic harmonic oscillator centered at some point different from the coordinate origin, but the multiplier $\mu$ depends on the distance to the origin, and not to the potential center, so after multiplying by $\mu$ this identification is no longer true. 

First, the presence of the additional terms in the potential breaks the rotational invariance, and prevents the conservation of the angular momentum, but the dynamics admits as  integral of motion  just a linear combination of the three components $J_i$, $i=1,2,3$, of the angular momentum with the constants $k_i$, $i=2,3,4$,  as coefficients 
\begin{equation}
 I  = k_2 J_{1} + k_3 J_{2} + k_4 J_{3}  \,,{\qquad} 
\bigl\{ I \,, H_\mu\bigr\}  =0 \,. 
\end{equation}

The factor $\mu$ preserves the Hamilton-Jacobi separability in Cartesian coordinates;
therefore we obtain the following three $\la$-dependent  integrals of motion 
$$
 K_{xx\la} = p_x^2 +  2(k_1x^2 + k_2 x) + 2 \la x^2 H_\mu  \,,{\quad} 
 K_{yy\la} = p_y^2 + 2(k_1y^2 + k_3 y) + 2 \la y^2 H_\mu  \,,{\quad} 
$$
\begin{equation}
 K_{zz\la} = p_z^2 + 2(k_1z^2 + k_4 z) + 2 \la z^2 H_\mu \,. 
\end{equation}
They are independent, that is $dK_{xx\la}\,\wedge\, dK_{yy\la}\,\wedge\, dK_{zz\la}\ne 0$,
 and satisfy 
$$ K_{xx\la} + K_{yy\la} + K_{zz\la} = 2 H_\mu \,,{\quad} 
\bigl\{ K_{aa\la}\,, H_{\mu}\bigr\} =  0 \,,{\quad} 
\bigl\{ K_{aa\la}\,, K_{bb\la}\bigr\} =  0 \,,{\quad} a,b =x,y,z. 
$$
In addition there is another set of three quadratic integrals $K_{ab\la}$, rather similar to the above three functions $K_{aa\la}$, that have the form  
$$
K_{xy\la} = p_x p_y + (2 k_1 x y + k_3 x + k_2 y) + 2 \la x y H_\mu  \,,{\quad} 
K_{yz\la} = p_y p_z + (2 k_1 y z +k_4 y + k_3 z) + 2 \la y z H_\mu  \,,
$$
\begin{equation}
 K_{zx\la} = p_z p_x + (2 k_1 z x +k_4 x + k_2 z) + 2 \la z x H_\mu  \,. 
\end{equation}
In fact, all of them can be grouped in a symmetric matrix $[K_{ab\la}]$
$$ M_K =\left[ K_{ab\la} \right]  =
\left[\matrix{
 K_{xx\la}  & K_{xy\la}  & K_{zx\la} \cr
 K_{xy\la}  & K_{yy\la}  & K_{yz\la} \cr 
 K_{zx\la}  & K_{yz\la}  & K_{zz\la} \cr  }\right],
$$
and we can summarize all the Poisson brackets with the  Hamiltonian in a single equation 
$$
 \Bigl\{K_{ab\la}\,,H_\mu\Bigr\} = 0 \,,{\quad} a,b =x,y,z. 
$$
The properties of $\left[ K_{ab\la} \right]$, that represents the $\la$-dependent version of the Fradkin tensor \cite{Frad65}, are summarized in the Appendix.

All these results are  summarized in the following proposition: 
\begin{proposition} \label{Prop1}
The $\la$-dependent Hamiltonian with a conformally Euclidean metric
$$
 H_\mu = \mu H_{111}  =   ({\smallonehalf})\,\Bigl( \frac{p_x^2 + p_y^2 +p_z^2}{1 - \la\,r^2} \Bigr)  
 +  \Bigl[\,  k_1\frac{x^2+y^2+z^2}{1 - \la\,r^2} + \frac{k_2\, x }{1 - \la\,r^2} +  \frac{k_3\, y}{1 - \la\,r^2} + \frac{k_4\, z }{1 - \la\,r^2}  \,\Bigr]     
$$
is superintegrable with five  functionally independent quadratic in the momenta constants of motion  in a family of six $\la$-dependent functions $K_{ab\la}$, $a,b =x,y,z$.   Three of them,  $K_{xx\la}$, $K_{yy\la}$, and $K_{zz\la}$,  Poisson commute among them and two other functions  in the set $K_{ab\la}$, $a\ne b$, can be chosen for the total set of five functionally independent integrals of motion. 
\end{proposition}

We close the study of this system by considering some geometric properties of the associated metric of the system. 
First we recall that the metric determines the kinetic term of the Lagrangian (coefficient of the Lagrangian of the geodesic motion) and that in this case is a conformally flat metric 
\begin{equation}
g_{ij} = (1 -\la\,r^2){\rm diagonal}[1,1,1]  \,,{\quad} r^2= x^2+y^2+z^2 \,. 
\end{equation}
The sectional curvatures with respect to the planes  $(x,y)$, $(x,z)$, and  $(y,z)$, that we denote by $\kappa_{xy}$, $\kappa_{xz}$,  and $\kappa_{yz}$,   take  respectively the forms
\begin{equation}
  \kappa_{xy} = \la\Bigl(\frac{2 - 3 \la z^2}{(1- \la r^2)^3} \Bigr)   \,,{\qquad}
  \kappa_{xz} = \la\Bigl(\frac{2 - 3 \la y^2}{(1- \la r^2)^3} \Bigr)   \,,{\qquad}
  \kappa_{yz} = \la\Bigl(\frac{2 - 3 \la x^2}{(1- \la r^2)^3} \Bigr)   \,, 
\end{equation}
and then the scalar curvature (Ricci curvature), that is given by two times the sum of the three sectional curvatures (it represents a mean value of the other partial curvatures), is 
\begin{equation}
 R = 6 \la\Bigl(\frac{2 -  \la r^2}{(1- \la r^2)^3}\Bigr) \,.
\end{equation}
We recall that the term $(1- \la r^2)$  is always positive and therefore both numerator and denominator in $R$ are positive. 
Thus, as $R$ has $\la$ as a global factor, then the curvature of the geometry is definite positive or negative according to the sign of $\la$.

\subsection{Isotropic oscillator with nonlinear terms $k_2/x^2$, $k_3/y^2$, and $k_4/z^2$} \label{Sec22}

Let us now denote by  $H_{111}$ the Hamiltonian  of the three-dimensional isotropic oscillator with additional nonlinear terms of the form $k_2/x^2$, $k_3/y^2$, and $k_4/z^2$
\begin{equation}
  H_{111} = ({\smallonehalf})\,\bigl( p_x^2 + p_y^2 +p_z^2 \bigr)   +  \Bigl[\, k_1(x^2+y^2+z^2) + \frac{k_2}{x^2} + \frac{k_3}{y^2} + \frac{k_4}{z^2}  \,\Bigr]     \,, 
\end{equation}
that is the three-dimensional version of the two-dimensional Smorodinsky-Winternitz system \cite{FrMS65, Ev90PL} 
(this system, that is known as the `caged oscillator' \cite{MillPW13, VeEv08b, KaMil12, ChanuLuRas15,GubLat18}, can be considered as the three-dimensional version of the isotonic oscillator).
It admits separability in several coordinate systems \cite{Ev90PR}, and it  is  superintegrable with quadratic constants of motion.

Now let us denote by  $\mu$ the following multiplier
\begin{equation}
 \mu  = 1/(1 -\la\,r^2) \,,{\quad} r^2= x^2+y^2+z^2 \,, 
\end{equation}
where $\la$ is a real parameter. Then the new $\la$-dependent Hamiltonian $H_\mu$ defined as 
$$
 H_{\mu} = \mu H_{111} = \bigl(\frac{1}{1-\la\, r^2}\bigr) H_{111} \,,{\quad} 
 \lim{}_{\la\to 0}  H_{\mu} =  H_{111} \,, 
$$
takes the form 
\begin{equation}
 H_\mu  =  \mu H_{111} = ({\smallonehalf})\,\Bigl( \frac{p_x^2 + p_y^2 +p_z^2}{1 - \la\,r^2} \Bigr)  
 +  \Bigl[\, k_1\frac{x^2+y^2+z^2}{1 - \la\,r^2}  + \frac{k_2/x^2 }{1 - \la\,r^2} +  \frac{k_3/y^2}{1 - \la\,r^2} + \frac{k_4/z^2 }{1 - \la\,r^2}  \,\Bigr]     \,. 
\end{equation}

The factor $\mu$ preserves the Hamilton-Jacobi separability in Cartesian coordinates. 
Therefore we obtain the following three $\la$-dependent integrals of motion 
$$
K_{1\la} = p_x^2 + 2 \Bigl( k_1x^2 + \frac{k_2}{x^2}\Bigr) + 2 \la x^2 H_\mu  \,,{\quad} 
K_{2\la} = p_y^2 + 2 \Bigl( k_1y^2 + \frac{k_2}{x^2}\Bigr) + 2 \la y^2 H_\mu  \,,{\quad} 
$$
\begin{equation}
  K_{3\la} = p_z^2 + 2\Bigl( k_1z^2 + \frac{k_4}{z^2}\Bigr) + 2 \la z^2 H_\mu  \,. 
\end{equation}
They are   functionally independent, that is $dK_{1\la}\,\wedge\, dK_{2\la}\,\wedge\, dK_{3\la}\ne 0$,
 and satisfy 
$$ K_{1\la} + K_{2\la} + K_{3\la} = 2 H_\mu \,,{\quad} 
\bigl\{ K_{i\la}\,, H_{\mu}\bigr\} =  0 \,,{\quad}    
\bigl\{ K_{i\la}\,, K_{j\la}\bigr\} =  0 \,,{\quad} i,j=1,2,3. 
$$
There is a second set of integrals of motion related to the components of the angular momentum that are $\la$-independent
$$
 K_{J1} = (y p_z - z p_y)^2  +  2 k_3 \Bigl(\frac{z}{y}\Bigr)^2  +  2 k_4 \Bigl(\frac{y}{z}\Bigr)^2  \,,{\quad} 
 K_{J2} = (z p_x - x p_z)^2  +  2 k_2 \Bigl(\frac{z}{x}\Bigr)^2  +  2 k_4 \Bigl(\frac{x}{z}\Bigr)^2  \,, 
$$ 
\begin{equation}
  K_{J3}  = (x p_y - y p_x)^2  +  2 k_2 \Bigl(\frac{y}{x}\Bigr)^2  +  2 k_3 \Bigl(\frac{x}{y}\Bigr)^2  \,.
\end{equation}
and functionally independent 
$$
 dK_{J1}\,\wedge\, dK_{J2}\,\wedge\, dK_{J3}\ne 0 \,.
$$
That is, the  factor $\mu$ modifies the Hamiltonian (and the dynamical vector field $X_\mu$) but these three  integrals remain invariant. 
Two of these three functions can be chosen for the total set of five  functionally independent integrals of motion.

We summarize the results in the following proposition 
\begin{proposition} \label{Prop2}
The $\la$-dependent Hamiltonian with a conformally Euclidean metric
$$
 H_\mu  =  \mu H_{111} =   ({\smallonehalf})\,\Bigl( \frac{p_x^2 + p_y^2 +p_z^2}{1 - \la\,r^2} \Bigr)  
 +  \Bigl[\, k_1\frac{x^2+y^2+z^2}{1 - \la\,r^2}  + \frac{k_2/x^2 }{1 - \la\,r^2} +  \frac{k_3/y^2}{1 - \la\,r^2} + \frac{k_4/z^2 }{1 - \la\,r^2}  \,\Bigr]     
$$
is superintegrable with two sets of three quadratic integrals of motion. A first set of three $\la$-dependent functions $K_{i\la}$, $i=1,2,3$, that Poisson commute and a second set  of three angular  momentum-related functions $K_{Ji}$,  $i=1,2,3$, that are $\la$-independent. 
Two of the functions in the second set  can be chosen for the total set of five functionally independent integrals of motion. 
\end{proposition}

\section{Oscillator  1:1:2 related Hamiltonian with nonlinear terms $k_2/x^2$, $k_3/y^2$, and $k_4/z^2$}  \label{Sec3}

\subsection{Oscillator 1:1:2 related Hamiltonian with Euclidean metric} \label{Sec31}

Let us consider the Hamiltonian  $H_{k23}$ of the harmonic oscillator with ratio of frequencies 1:1:2 and with two additional nonlinear terms of the form  $k_2/x^2$ and $k_3/y^2$  
\begin{equation}
 H_{k23}  =  ({\smallonehalf})\,\bigl( p_x^2 + p_y^2 +p_z^2 \bigr)  
 +  \Bigl[  k_1 (x^2+y^2+4 z^2) + \frac{k_2 }{x^2} +  \frac{k_3 }{y^2}   \,\Bigr]     \,.  
\end{equation}
It is Hamilton-Jacobi separable in two different systems of coordinates, Cartesian $(x,y,z)$ and parabolic $(\sigma,\tau,\phi)$ \cite{Ev90PR}, and it is therefore superintegrable with five  functionally independent quadratic constants of motion. 

\begin{itemize}
\item[(i)] Three functions arise from the separability of the Hamilton-Jacobi equation in Cartesian coordinates
\begin{equation}
 K_1 =   p_x^2 +  2 k_1 x^2 + 2 k_2/x^2  \,,{\quad} 
 K_2 =   p_y^2 +  2 k_1 y^2 + 2 k_3/y^2  \,,{\quad} 
 K_3 =   p_z^2 +  8 k_1  z^2  \,,
\end{equation}
$$
 K_1 +K_2 +K_3 = 2 H_{k23} \,,{\quad}  \bigl\{ K_{i}\,, H_{k23}\bigr\}  = 0 \,,{\quad}  
 \bigl\{ K_{i}\,, K_{j}\bigr\} =  0 \,,{\quad} i,j = 1,2,3. 
$$
\item[(ii)] A constant of motion related with the angular momentum 
\begin{equation}
  K_4 = K_{J3}  = J_{3}^2  +  2 k_2 \Bigl(\frac{y}{x}\Bigr)^2  +  2 k_3 \Bigl(\frac{x}{y}\Bigr)^2   \,,{\quad} 
\bigl\{ K_{J3}\,, H_{k23}\bigr\}  = 0 \,. 
\end{equation}
\item[(iii)] Two constants of motion of Runge-Lenz type  structure 
\begin{eqnarray}
 K_{RL1} &=&  - p_x J_2 + 2 k_1 x^2  z  - 2 k_2 (z/x^2) 
\,,{\quad}  \bigl\{ K_{RL1}\,, H_{k23}\bigr\}  = 0 \,, \cr
 K_{RL2} &=&  p_y J_1  + 2 k_1y^2  z  -  2 k_3 (z/y^2)  
\,,{\quad}  \bigl\{ K_{RL2}\,, H_ {k23}\bigr\}  = 0 \,, 
\end{eqnarray}
that are functionally independent, that is $dK_{RL1}\,\wedge\, dK_{RL2}\ne 0$.
One of these two Runge-Lenz functions, $K_{RL1}$ or $K_{RL2}$, can be chosen for the total set of five independents integrals of motion. 
\end{itemize}

But our purpose is the study of the following more general Hamiltonian \cite{VeEv08b}
\begin{equation}
 H_{k234}  =  ({\smallonehalf})\,\bigl( p_x^2 + p_y^2 +p_z^2 \bigr)  +  
 \Bigl[\,  k_1 (x^2+y^2+4 z^2) + \frac{k_2 }{x^2} +  \frac{k_3 }{y^2} +  \frac{k_4 }{z^2} \,\Bigr]   \,.   
\end{equation}
The new term $k_4/z^2$ destroys the separability in parabolic coordinates and therefore the fifth and the sixth constants of motion,  $K_{RL1}$ and $K_{RL2}$, are not preserved (that is, $\{ K_{RL1}\,, H_{k234}\}  \ne 0$ and $\{ K_{RL2}\,, H_{k234}\}\ne 0$). 
The consequence is that this new system has only four quadratic first integrals (we recall that a 3-dimensional system is called minimally superintegrable if it admits only four  functionally independent globally defined integrals of motion). 

Next we will prove that actually $H_{k234}$ is  maximally superintegrable. 
The important point is that it admits new constants of motion but of higher order. 
That is, the new nonlinear term $k_4/z^2$ prevents the existence of the above mentioned quadratic integrals of motion, $K_{RL1}$ and $K_{RL2}$,  but it gives rise to the existence of several constants of motion  of fourth order,  all of them not arising from separability.

The proof is as follows. The  two pair of functions 
\begin{equation}
 \Bigl( K_{RL1} \,,\,\frac{1}{z} \bigl(x p_x  \bigr)\Bigr) 
\quad{\rm and}{\quad}
 \Bigl( K_{RL2} \,,\, \frac{1}{z} \bigl(y p_y  \bigr)\Bigr) 
\end{equation}
are related between them by the time derivatives. 
More precisely we have
$$
  \frac{d}{dt}\,K_{RL1} = 2 k_4 \lambda_z\,\Bigl[\frac{1}{z} \bigl(x p_x \bigr)\Bigr] \,,{\qquad}
  \frac{d}{dt}\,\Bigl[\frac{1}{z} \bigl(x p_x  \bigr) \Bigr] =  -\, \lambda_z\,K_{RL1} \,,{\qquad}
  \lambda_z = \frac{1}{z^2} \,,
$$
and
$$
  \frac{d}{dt}\,K_{RL2} = 2 k_4 \lambda_z\,\Bigl[\frac{1}{z} \bigl(y p_y \bigr)\Bigr] \,,{\qquad}
  \frac{d}{dt}\,\Bigl[\frac{1}{z} \bigl(y p_y  \bigr) \Bigr] = -\, \lambda_z\,K_{RL2} \,,{\qquad}
  \lambda_z = \frac{1}{z^2} \,,
$$
where of course the time-derivative means Poisson bracket with the Hamiltonian $H_{k234} $.
Therefore, if we denote by $M_1$ and $M_2$ the following complex function
$$
 M_1 = K_{RL1} + {\ii} \sqrt{2 k_4}\,\,\Bigl[\frac{1}{z} \bigl(x p_x  \bigr)\Bigr] \,,
 {\qquad}
 M_2 = K_{RL2} + {\ii} \sqrt{2 k_4}\,\,\Bigl[\frac{1}{z} \bigl(y p_y  \bigr)\Bigr] \,,
$$
then we have 
$$
 \frac{d}{dt}\,M_1 = \frac{d}{dt}\,K_{RL1} + 
 {\ii} \sqrt{2 k_4}\,\,\frac{d}{dt}\,\Bigl[\frac{1}{z} \bigl(x p_x  \bigr) \Bigr] 
 = -\, {\ii}\sqrt{2 k_4}\, \lambda_z\,M_1 \,, 
 $$
$$
 \frac{d}{dt}\,M_2 = \frac{d}{dt}\,K_{RL2} + 
 {\ii} \sqrt{2 k_4}\,\,\frac{d}{dt}\,\Bigl[\frac{1}{z} \bigl(y p_y  \bigr) \Bigr] 
 = -\, {\ii}\sqrt{2 k_4}\, \lambda_z\,M_2 \,, 
 $$
 and consequently 
\begin{eqnarray*}
 \bigl\{ M_1 M_1^*\,, H_{k234}\bigr\}  
  &=& \bigl\{ M_1 \,, H_{k234}\bigr\} M_1^* + M_1  \bigl\{ M_1^*\,, H_{k234}\bigr\}   \cr
  &=& \Bigl(-\, {\ii} \sqrt{2 k_4}\, \lambda_z M_1 \Bigr) M_1^*+   M_1\Bigl( {\ii}\sqrt{2 k_4}\, \lambda_zM_1^*  \Bigr)= 0 \,,  
\end{eqnarray*}
\begin{eqnarray*}
 \bigl\{ M_2 M_2^*\,, H_{k234}\bigr\}  
  &=& \bigl\{ M_2 \,, H_{k234}\bigr\} M_2^* + M_2  \bigl\{ M_2^*\,, H_{k234}\bigr\}   \cr
  &=& \Bigl(-\, {\ii} \sqrt{2 k_4}\, \lambda_z M_2 \Bigr) M_2^*+   M_2\Bigl( {\ii}\sqrt{2 k_4}\, \lambda_zM_2^*  \Bigr)= 0 \,. 
\end{eqnarray*}
\begin{eqnarray*}
 \bigl\{ M_1 M_2^*\,, H_{k234}\bigr\}  
  &=& \bigl\{ M_1 \,, H_{k234}\bigr\} M_2^* + M_1  \bigl\{ M_2^*\,, H_{k234}\bigr\}   \cr
  &=& \Bigl(-\, {\ii} \sqrt{2 k_4}\, \lambda_z M_1 \Bigr) M_2^*+   M_1\Bigl( {\ii}\sqrt{2 k_4}\, \lambda_zM_2^*  \Bigr)= 0 \,. 
\end{eqnarray*}
Thus the following four functions 
\begin{equation}
  K_{5} ={\rm Im}(M_1 M_2^*)  \,,{\quad} K_{6a} ={\rm Re}(M_1 M_2^*)  \,,{\quad}
  K_{6b} =  | \, M_1\, |^2 \,,{\quad} K_{6c} =  | \, M_2 \,|^2  \,,
\end{equation}
are all of them integrals of motion. The function $K_{5}$ is cubic in the momenta 
\begin{equation}
 K_{5}   =  \frac{\sqrt{2 k_4}}{z} \bigl(y p_y  \bigr)\bigl(K_{RL1}\bigr) -  \frac{\sqrt{2 k_4}}{z} \bigl(x p_x  \bigr)\bigl(K_{RL2}\bigr)\,,{\quad} 
  \bigl\{ K_{5}\,, H_{k234}\bigr\}  = 0 \,,  
 \end{equation}
but it is not functionally  independent of the others (that is, $dK_{1}\,\wedge\, dK_{2}\,\wedge\, dK_{3}\,\wedge\, dK_{4}\,\wedge\, dK_{5}= 0$).  
The other three, $K_{6a}$, $K_{6b}$, and $K_{6c}$, are of the fourth order in the momenta
\begin{eqnarray}
  K_{6a}   &=& \bigl(K_{RL1}\bigr)\bigl(K_{RL2}\bigr) + \frac{2k_4}{z^2} \bigl(x p_x  \bigr) \bigl(y p_y  \bigr)\,,{\quad} 
  \bigl\{ K_{6a}\,, H_{k234}\bigr\}  = 0 \,. \cr
 K_{6b}   &=& \bigl(K_{RL1}\bigr)^{2} + \frac{2k_4}{z^2} \bigl(x p_x  \bigr)^2 \,,{\quad} 
  \bigl\{ K_{6b}\,, H_{k234}\bigr\}  = 0 \,, \cr
 K_{6c}   &=& \bigl(K_{RL2}\bigr)^{2} + \frac{2k_4}{z^2} \bigl(y p_y \bigr)^2 \,,{\quad} 
  \bigl\{ K_{6c}\,, H_{k234}\bigr\}  = 0 \,. 
\end{eqnarray}
Next we note two properties. First,  the two  fourth-order functions $K_{6b} $ and $K_{6c} $ satisfy the following limit 
$$
 \lim{}_{k_4\to 0}K_{6b} =  \bigl(K_{RL1}\bigr)^{2}  \,,
  {\quad}
 \lim{}_{k_4\to 0}K_{6c} =  \bigl(K_{RL2}\bigr)^{2}  \,, 
$$
so that we recover the  Runge-Lenz like quadratic constants of motion $K_{RL1}$ and $K_{RL2}$ of the Hamiltonian $H_{k23}$. 
Second, these functions can be grouped in a symmetric matrix $[M_{K6}]$ whose determinant states the functional relation of  the quartic functions with the cubic function $K_5$  
$$ \left[ M_{K6} \right]  =
\left[\matrix{
K_{6b}  &  K_{6a}  \cr
K_{6a}  &  K_{6c}  \cr  }\right]  \,,{\quad}
 \det\left[\, M_{K6} \right]  = (K_5)^2 \,.
$$
We can summarize these results as follows:

\begin{proposition}  \label{Prop3}
 The   Hamiltonian of the 1:1:2  oscillator with three nonlinear terms 
$$
 H_{k234}  =  ({\smallonehalf})\,\bigl( p_x^2 + p_y^2 +p_z^2 \bigr)  
 +  \Bigl[\,  k_1 (x^2+y^2+4 z^2) + \frac{k_2 }{x^2} +  \frac{k_3 }{y^2} +  \frac{k_4 }{z^2}   \,\Bigr]      
$$
is maximally superintegrable with a fundamental set of three quadratic constants of motion $(K_{1}, K_{2}, K_{3})$ that Poisson commute, a fourth quadratic first  integral related to the angular momentum $(K_4=K_{J3})$ and three additional constants of motion, $K_{6a}$, $K_{6b}$  and $K_{6c}$, of fourth order in the momenta. 
One of these three quartic functions can be chosen for the total set of five functionally independents integrals of motion. 
\end{proposition}

\subsection{Oscillator 1:1:2 related Hamiltonian with a conformally Euclidean metric} 

Now let us denote by  $\mu$ the following multiplier
\begin{equation}
 \mu  = 1/(1 -\la\,f) \,,{\quad} f=(x^2+y^2+ 4z^2)  \,, 
\end{equation}
where $\la$ is a real parameter defined is such a way that $\mu$ must be positive, i.e. when $\lambda>0$ the configuration space is restricted by $x^2+y^2+ 4z^2<1/\lambda$. 
The new Lagrangian is 
\begin{equation}
 L_{\mu}  =   ({\smallonehalf})\,\bigl(1 - \la\, (x^2+y^2+ 4z^2)\bigr)\bigl( v_x^2 + v_y^2 + v_z^2 \bigr)  
 -   \Bigl[\,  \frac{k_1 (x^2+y^2+ 4z^2)}{1 - \la\, (x^2+y^2+ 4z^2)} + 
 \frac{k_2/x^2 + k_3/y^2+ k_4/z^2}{1 - \la\, (x^2+y^2+ 4z^2)}  \,\Bigr]    \,, 
\end{equation}
and the new Hamiltonian, that is given by 
 $$
 H_{\mu} = \mu H_{k234} = \Bigl(\frac{1}{1 - \la\, (x^2+y^2+ 4z^2)}\Bigr) H_{k234} \,,{\quad} 
 \lim{}_{\la\to 0}  H_{\mu} =  H_{k234}  \,, 
$$
takes the form 
\begin{equation}
 H_\mu  =   ({\smallonehalf})\,\Bigl( \frac{p_x^2 + p_y^2 +p_z^2}{1 - \la\, (x^2+y^2+ 4z^2)} \Bigr)  
 +  \Bigl[\,  \frac{k_1 (x^2+y^2+ 4z^2)}{1 - \la\, (x^2+y^2+ 4z^2)} + 
 \frac{k_2/x^2 + k_3/y^2+ k_4/z^2}{1 - \la\, (x^2+y^2+ 4z^2)}  \,\Bigr]     \,. 
\end{equation}
\begin{itemize}
\item[(i)]  The factor $\mu$ preserves the separability in Cartesian coordinates; 
 consequently   the following three $\la$-dependent functions are integrals of motion:
$$
 K_{1\la} =   p_x^2 +  2  k_1 x^2 +  2 k_2/x^2  +  2 \la x^2  H_\mu  \,,{\quad} 
 K_{2\la} =   p_y^2 +  2  k_1 y^2 +  2 k_3/y^2  +  2 \la y^2  H_\mu  \,,
$$ 
\begin{equation}
 K_{3\la} =   p_z^2 +  8  k_1 z^2 +  2 k_4/z^2  +  8 \la z^2  H_\mu   \,,
\end{equation}
that are  functionally independent, that is $dK_{1\la}\,\wedge\, dK_{2\la}\,\wedge\, dK_{3\la}\ne 0$, and satisfy 
$$ 
K_{1\la} + K_{2\la} + K_{3\la} = 2 H_\mu \,,{\quad} 
\bigl\{ K_{i\la}\,, H_{\mu}\bigr\} =  0 \,,{\quad}
\bigl\{ K_{i\la}\,, K_{j\la} \bigr\} =  0 \,,{\quad}  i,j=1,2,3.
$$
\item[(ii)] The factor $\mu$ preserves the expression of the fourth integral of motion that is  $\la$-independent 
\begin{equation}
  K_4 = K_{J3}  = (xp_y-yp_x)^2  +  2 k_2 \Bigl(\frac{y}{x}\Bigr)^2  +  2 k_3 \Bigl(\frac{x}{y}\Bigr)^2  
   \,,{\quad}   \bigl\{ K_{J3}\,, H_\mu\bigr\}  = 0 \,. 
\end{equation}
\end{itemize}

Now if we denote by $K_{RL1\mu}$ and $K_{RL2\mu}$ he following $\la$-dependent functions obtained as a deformation of the previous Runge-Lenz like functions $K_{RL1}$ and $K_{RL2}$ 
$$
  K_{RL1\mu} = K_{RL1} + 2\,\la  \mu x^2 z H_{k234} \,,{\quad} 
  \lim{}_{\la\to 0}  K_{RL1\mu} =  K_{RL1}  \,,
$$
$$
  K_{RL2\mu} = K_{RL2} + 2\,\la  \mu y^2 z H_{k234} \,,{\quad} 
  \lim{}_{\la\to 0}  K_{RL2\mu}=  K_{RL2}  \,, 
$$
then we have the following two properties 
$$
  \frac{d}{dt}\,K_{RL1\mu} = 2 k_4 \lambda_\mu\,\Bigl[\frac{1}{z} \bigl(x p_x \bigr)\Bigr] \,,{\quad}
  \frac{d}{dt}\,\Bigl[\frac{1}{z} \bigl(x p_x  \bigr) \Bigr] 
  = -\, \lambda_\mu\,K_{RL1\mu} \,,{\quad}  \lambda_\mu = \mu\,\frac{1}{z^2} \,,  
$$
$$
  \frac{d}{dt}\,K_{RL2\mu} = 2 k_4 \lambda_\mu\,\Bigl[\frac{1}{z} \bigl(y p_y \bigr)\Bigr] \,,{\quad}
  \frac{d}{dt}\,\Bigl[\frac{1}{z} \bigl(y p_y  \bigr) \Bigr] 
  = -\, \lambda_\mu\,K_{RL2\mu} \,,{\quad}  \lambda_\mu = \mu\,\frac{1}{z^2} \,. 
$$
They are rather similar to the previous properties in the case of the Euclidean Hamiltonian $H_{k234}$ but with $\la$-dependent functions; 
that is, $H_{\mu}$ instead of $H_{k234}$, $K_{RL1\mu}$ and $K_{RL2\mu}$ instead of $K_{RL1}$ and $K_{RL2}$ and $\lambda_\mu$ instead of $\lambda_z$. 
The complex functions $M_1$ and $M_2$ of the Euclidean case  are now the following $\la$-dependent complex functions  $M_{1\mu}$ and $M_{2\mu}$ : 
$$
 M_{1\mu} = K_{RL1\mu} + {\ii} \sqrt{2 k_4}\,\,\Bigl[\frac{1}{z} \bigl(x p_x \bigr)\Bigr] \,,
{\qquad}
 M_{2\mu} = K_{RL2\mu} + {\ii} \sqrt{2 k_4}\,\,\Bigl[\frac{1}{z} \bigl(y p_y \bigr)\Bigr] \,,
$$
that satisfy
$$
  \bigl\{ M_{1\mu}\,, H_{\mu}\bigr\}  = -\, {\ii}\sqrt{2 k_4}\, \lambda_{\mu}\,M_{1\mu}\,,
{\qquad}
  \bigl\{ M_{2\mu}\,, H_{\mu}\bigr\}  = -\, {\ii}\sqrt{2 k_4}\, \lambda_{\mu}\,M_{2\mu}\,,
$$
and therefore we obtain the following results 
$$
 \bigl\{ M_{1\mu} M_{1\mu}^*\,, H_{\mu}\bigr\}   
 = \Bigl(-\, {\ii} \sqrt{2 k_4}\, \lambda_{\mu} M_{1\mu} \Bigr) M_{1\mu}^* 
 +   M_{1\mu}\Bigl( {\ii}\sqrt{2 k_4}\, \lambda_{\mu}M_{1\mu}^*  \Bigr)= 0 \,, 
$$
$$
 \bigl\{ M_{2\mu} M_{2\mu}^*\,, H_{\mu}\bigr\}   
  = \Bigl(-\, {\ii} \sqrt{2 k_4}\, \lambda_{\mu} M_{2\mu} \Bigr) M_{2\mu}^* 
  +   M_{2\mu}\Bigl( {\ii}\sqrt{2 k_4}\, \lambda_{\mu}M_{2\mu}^*  \Bigr)= 0 \,,  
$$
$$
 \bigl\{ M_{1\mu} M_{2\mu}^*\,, H_{\mu}\bigr\}  
   = \Bigl(-\, {\ii} \sqrt{2 k_4}\, \lambda_z M_{1\mu}  \Bigr) M_{2\mu}^* 
   +  M_{1\mu}\Bigl( {\ii}\sqrt{2 k_4}\, \lambda_zM_{2\mu}^*  \Bigr)= 0 \,. 
$$
Thus the following four $\la$-dependent functions
\begin{equation}
  K_{5\la} ={\rm Im}(M_{1\mu} M_{2\mu}^*)  \,,{\quad} 
  K_{6a\la} ={\rm Re}(M_{1\mu} M_{2\mu}^*)  \,,{\quad}
  K_{6b\la} =  | \, M_{1\mu}\, |^2 \,,{\quad} K_{6c\la} =  | \, M_{2\mu} \,|^2  \,,
\end{equation}
are all of them integrals of motion. 
As in the Euclidean case, one of them is cubic and the other three are quartic.
That is, the  function $K_{5\la}$  given by 
\begin{equation}
 K_{5\la}   =  \frac{\sqrt{2 k_4}}{z} \bigl(y p_y  \bigr)\bigl(K_{RL1\mu}\bigr) -   
 \frac{\sqrt{2 k_4}}{z} \bigl(x p_x  \bigr)\bigl(K_{RL2\mu}\bigr)\,,{\quad} 
  \bigl\{ K_{5\la}\,, H_{\mu}\bigr\}  = 0 \,,  
\end{equation}
is of third order in the momenta but it is functionally dependent of the four quadratic first integrals. 
The other three, $K_{6a\la}$, $K_{6b\la}$, and $K_{6c\la}$, are of the fourth order in the momenta
\begin{eqnarray}
 K_{6a\la}   &=& \bigl(K_{RL1\mu}\bigr)\bigl(K_{RL2\mu}\bigr) + \frac{2k_4}{z^2} \bigl(x p_x  \bigr) \bigl(y p_y  \bigr)\,,{\quad} 
  \bigl\{ K_{6a\la}\,, H_{\mu}\bigr\}  = 0 \,, \cr
 K_{6b\la}   &=& \bigl(K_{RL1\mu}\bigr)^{2} + \frac{2k_4}{z^2} \bigl(x p_x  \bigr)^2 \,,{\quad} 
  \bigl\{ K_{6b\la}\,, H_{\mu}\bigr\}  = 0 \,, \cr
 K_{6c\la}  &=& \bigl(K_{RL2\mu}\bigr)^{2} + \frac{2k_4}{z^2} \bigl(y p_y \bigr)^2 \,,{\quad} 
  \bigl\{ K_{6c\la}\,, H_{\mu}\bigr\}  = 0  \,. 
\end{eqnarray}
The deformation introduced by the parameter $\la$ preserves the relation obtained in the Euclidean case between the  quartic and the cubic functions; that is, we have a $\la$-dependent symmetric matrix $[M_{K6\la}]$ and the determinant of this matrix is just the square of the cubic function  
$$ \left[ M_{K6\la} \right]  =
\left[\matrix{
K_{6b\la}  &  K_{6a\la}  \cr
K_{6a\la}  &  K_{6c\la}  \cr  }\right]  \,,{\quad}
 \det\left[\, M_{K6\la} \right]  = (K_{5\la})^2 \,.
$$

The following proposition summarizes the results we have obtained:
\begin{proposition} \label{Prop4}
The $\la$-dependent Hamiltonian with a conformally Euclidean metric
$$
 H_\mu = \mu H_{k234} = \Bigl(\frac{1}{1 - \la\, (x^2+y^2+ 4z^2)}\Bigr) H_{k234}
  \,,{\quad} 
 \lim{}_{\la\to 0}  H_{\mu} =  H_{k234} \,, 
$$
where $H_{k234}$ is the Hamiltonian of the 1:1:2 oscillator with three nonlinear terms 
$$
 H_{k234}  =  ({\smallonehalf})\,\bigl( p_x^2 + p_y^2 +p_z^2 \bigr)  +  
 \Bigl[\,  k_1 (x^2+y^2+4 z^2) + \frac{k_2 }{x^2} +  \frac{k_3 }{y^2} +  \frac{k_4 }{z^2} \,\Bigr]   \,, 
$$
is maximally superintegrable with a fundamental set of three $\la$-dependent quadratic constants of motion $(K_{1\la}, K_{2\la}, K_{3\la})$ that Poisson commute, a fourth quadratic first  integral related with the angular momentum $(K_4=K_{J3})$ and 
a set of three additional $\la$-dependent constants of motion, $K_{6a\la}$, $K_{6b\la}$  and $K_{6c\la}$, of fourth order in the momenta. 
One of these three quartic functions can be chosen for the total set of five functionally independents integrals of motion. 
\end{proposition}

In this case the conformally flat metric $g_{ij} $ is given by the metric in the kinetic term in the Lagrangian $L_\mu$ 
\begin{equation}
 g_{ij} = \bigl(1 - \la\, (x^2+y^2+ 4z^2)\bigr)\,{\rm diagonal}[1,1,1]
\end{equation}
and the sectional curvatures $\kappa_{xy}$, $\kappa_{xz}$,  and $\kappa_{yz}$,  with respect to the planes  $(x,y)$, $(x,z)$, and  $(y,z)$  take respectively the forms 
$$
\kappa_{xy} = 2\la\,\Bigl(\frac{ 1 - 12 \la z^2}{(1 - \la\, (x^2+y^2+ 4z^2))^3} \Bigr)  \,,{\qquad}
\kappa_{xz} = \la\,\Bigl(\frac{5 - 3 \la (x^2 + 2y^2 -4z^2)}{(1 - \la\, (x^2+y^2+ 4z^2))^3} \Bigr)  \,,
$$
\begin{equation}
\kappa_{yz} = \la\,\Bigl(\frac{5 - 3 \la (2x^2 + y^2 -4z^2)}{(1 - \la\, (x^2+y^2+ 4z^2))^3} \Bigr)     \,.
\end{equation}
The  expressions of these sectional curvatures are not very simple. 
They have $\la$ as a  global factor but the numerators also depend of $\la$ and can take positive or negative values depending of values of the coordinates (the denominators are positive since the dynamics is defined in the region  $(1 - \la\, (x^2+y^2+ 4z^2))>0$). The scalar curvature (Ricci curvature), that  is given by two times the sum of the three sectional curvatures, is
\begin{equation}
 R = 6 \la\,\Bigl(\frac{ 4- 3 \la(x^2+y^2)}{(1 - \la\, (x^2+y^2+ 4z^2))^3} \Bigr)\,,
\end{equation}
where the expression in parenthesis is always positive; therefore the sign of $R$ and  the type of geometry (negative hyperbolic or positive spherical) depends directly  on the value of  $\la$.

\section{Kepler related Hamiltonian with nonlinear terms $k_2/x^2$, $k_3/y^2$, and $k_4/z^2$}   \label{Sec4}

\subsection{Euclidean Kepler related Hamiltonian} \label{Sec41}

Let us consider the Hamiltonian  $H_{K234}$ of the Kepler problem with three additional nonlinear terms of the form  $k_2/x^2$, $k_3/y^2$ and $k_4/z^2$
\begin{equation}
  H_{K234}  =  ({\smallonehalf})\,\bigl( p_x^2 + p_y^2 +p_z^2 \bigr) 
  + V_{K234}
\,,   {\quad}
  V_{K234}  =  \frac{k_1}{r} + \frac{k_2}{x^2} + \frac{k_3}{y^2} + \frac{k_4}{z^2} \,.
\end{equation}
It is separable in spherical $(r,\te,\phi)$ coordinates and it is therefore Liouville integrable with quadratic integrals of motion.  In the particular case $k_4=0$ it is also separable in parabolic $(\sigma,\tau,\phi)$ coordinates and in this case  is superintegrable with five  functionally independent quadratic first integrals (including the Hamiltonian itself) \cite{Ev90PR}. Verrier {\sl et al}  \cite{VeEv08a}  and Rodriguez {\sl et al} \cite{RodTempW08PRe,RodTempW09JPCS}, proved, by making use of dimensional reduction and action-angle variables,  that in the general case $k_i\ne 0$, $i=2,3,4$,  it admits a fifth   first  integral quartic in the momenta (not arising from separability of the Hamilton-Jacobi equation)  so that the system is maximally superintegrable 
(the quadratic algebra of symmetries of this system was studied in  \cite{TaDas11}).

Now we prove the existence of a total of six integrals of motion (seven with the Hamiltonian but of course only five of them are  functionally independent) that can be grouped in two sets with  three integrals in each one.

First, the components $(J_1,J_2,J_3)$ of the angular momentum are not preserved but the following three angular momentum related functions
$$
  K_{J1}  = J_1^2  +  2 k_3 \Bigl(\frac{z}{y}\Bigr)^2  +  2 k_4 \Bigl(\frac{y}{z}\Bigr)^2  \,,{\quad} 
  K_{J2}  = J_2^2  +  2 k_2 \Bigl(\frac{z}{x}\Bigr)^2  +  2 k_4 \Bigl(\frac{x}{z}\Bigr)^2  \,,
$$
\begin{equation}
  K_{J3}  = J_3^2 +  2 k_2 \Bigl(\frac{y}{x}\Bigr)^2  +  2 k_3 \Bigl(\frac{x}{y}\Bigr)^2  \,, 
\end{equation}
are functionally independent, 
$ dK_{J1}\,\wedge\, dK_{J2}\,\wedge\, dK_{J3}\ne 0  $,
and  satisfy the following Poisson bracket properties  
$$
  \bigl\{ K_{J1}\,, K_{J2}+K_{J3}\bigr\} =  0\,,{\qquad}  
  \bigl\{ K_{Ji}\,, H_{K234}\bigr\} =  0 \,,{\qquad} i=1,2,3.
$$

Second, let us denote by $R_a$, $a=x,y,z$, the following Runge-Lenz-related functions 
\begin{eqnarray}
 R_{x} &=& (J_2 p_z - J_3 p_y) - x\,\Bigl(\frac{k_1}{r} + \frac{2k_2}{x^2} + \frac{2k_3}{y^2} + \frac{2k_4}{z^2} \Bigr)  \,, \cr 
 R_{y} &=& (J_3 p_x - J_1 p_z) - y\,\Bigl(\frac{k_1}{r} + \frac{2k_2}{x^2} + \frac{2k_3}{y^2} + \frac{2k_4}{z^2} \Bigr)  \,,  \cr
 R_{z} &=& (J_1 p_y - J_2 p_x) - z\,\Bigl(\frac{k_1}{r} + \frac{2k_2}{x^2} + \frac{2k_3}{y^2} + \frac{2k_4}{z^2} \Bigr)  \,.  
\end{eqnarray}
In fact, in the particular case $(k_1\ne 0,k_2=k_3=k_4=0)$, these three functions reduce to three components of the Runge-Lenz vector. 

We have the following property. The functions $R_{a}$, $a=x,y,z$, and the functions
$$ 
 \bigl(x p_x + y p_y + z p_z\bigr)/x \,,{\quad}  
 \bigl(x p_x + y p_y + z p_z\bigr)/y \,,{\quad}   
 \bigl(x p_x + y p_y + z p_z\bigr)/z \,,
$$
are related among them by the time derivatives. 
More precisely,  we have
$$
  \bigl\{ R_x\,, H_{K234} \bigr\} = 2 k_2 \lambda_x\,\frac{1}{x} \Bigl(x p_x + y p_y + z p_z\Bigr) \,,{\quad}
 \Bigl\{\frac{1}{x} \bigl(x p_x + y p_y + z p_z\bigr) \,, H_{K234} \Bigr\} 
 = -\, \lambda_x\,R_x \,, 
$$
$$
  \bigl\{R_y\,, H_{K234} \bigr\} = 2 k_3 \lambda_y\,\frac{1}{y} \Bigl(x p_x + y p_y + z p_z\Bigr ) \,,{\quad}
 \Bigl\{\frac{1}{y} \Bigl(x p_x + y p_y + z p_z\bigr) \,, H_{K234} \Bigr\}  
 =  -\, \lambda_y\,R_y \,,
$$
$$
   \bigl\{ R_z\,, H_{K234} \bigr\} = 2 k_4 \lambda_z\,\frac{1}{z} \Bigl(x p_x + y p_y + z p_z\Bigr) \,,{\quad}
 \Bigl\{\frac{1}{z} \Bigl(x p_x + y p_y + z p_z\bigr) \,, H_{K234} \Bigr\}  
 = -\, \lambda_z\,R_z \,,
$$
where the coefficients $\la_a$, $a=x,y,z$, take the forms
$$
  \lambda_x = \frac{1}{x^2} \,,{\quad}  \lambda_y = \frac{1}{y^2} \,,{\quad}
  \lambda_z = \frac{1}{z^2} \,. 
$$
Then the following proposition states the properties of these functions. 
\begin{proposition} \label{Prop5}
Let   $M_a$, $a=x,y,z$, denote  the following complex functions  
$$
 M_a = R_a + {\ii} \sqrt{2 k_j}\,\frac{1}{a} \Bigl(x p_x + y p_y + z p_z\Bigr) \,,{\quad}
 a = x,y,z,{\quad} j=2,3,4. 
$$
Then the time derivatives of everyone of these functions satisfy the following  relations
$$
  \frac{d}{dt}\,M_x = -\, {\ii} \sqrt{2 k_2}\, \lambda_x\,M_x \,,{\quad}
  \frac{d}{dt}\,M_y = -\, {\ii}\sqrt{2 k_3}\, \lambda_y\,M_y  \,,{\quad}
  \frac{d}{dt}\,M_z = -\, {\ii}\sqrt{2 k_4}\, \lambda_z\,M_z  \,.
$$
\end{proposition} 
Therefore the moduli  $|\, M_a\,|$ of the functions $M_a$, $a=x,y,z$, satisfy 
$$
  \frac{d}{dt}\,|\, M_x\,|^2 = \Bigl( \frac{d}{dt}\,M_x\Bigr) M_x^* +  \,M_x\Bigl( \frac{d}{dt}M_x^*\Bigr)  = (-\, {\ii} \sqrt{2 k_2}\, \lambda_x  + {\ii}\sqrt{2 k_2}\, \lambda_x )\Bigl(M_xM_x^*\Bigr) = 0 \,,
$$
$$
  \frac{d}{dt}\,|\,M_y\,|^2 = \Bigl( \frac{d}{dt}\,M_y\Bigr) M_y^* +  \,M_y\Bigl( \frac{d}{dt}M_y^*\Bigr)  = (-\, {\ii} \sqrt{2 k_3}\, \lambda_y  + {\ii} \sqrt{2 k_3}\, \lambda_y )\Bigl(M_yM_y^*\Bigr) = 0 \,,
$$
$$
  \frac{d}{dt}\,|\, M_z\,|^2 = \Bigl( \frac{d}{dt}\,M_z\Bigr) M_z^* +  \,M_z\Bigl( \frac{d}{dt}M_z^*\Bigr)  = (-\, {\ii} \sqrt{2 k_4}\, \lambda_z  + {\ii} \sqrt{2 k_4}\, \lambda_z )\Bigl(M_zM_z^*\Bigr) = 0 \,. 
$$
Hence the three functions  $K_{4a}$,  $a=x,y,z$, given by 
$$
 K_{4x} =|\, M_x\,|^2  = R_{x}^2 + \frac{2k_2}{x^2} \Bigl(x p_x + y p_y + z p_z\Bigr)^2  \,,{\quad}
 K_{4y} =|\, M_y\,|^2  = R_{y}^2 + \frac{2k_3}{y^2} \Bigl(x p_x + y p_y + z p_z\Bigr)^2 \,, 
$$
\begin{equation}
K_{4z} =|\,  M_z\,|^2   = R_{z}^2 + \frac{2k_4}{z^2} \Bigl(x p_x + y p_y + z p_z\Bigr)^2 \,, 
\end{equation}
are quartic  constants of motion of motion 
$$
  \bigl\{ K_{4a}\,, H_{K234}\bigr\} =  0 \,,{\qquad} a=x,y,z.
  $$

  We have proved   the following proposition: 
\begin{proposition} \label{Prop6}
 The Kepler Hamiltonian with three additional nonlinear terms 
$$
  H_{K234}  =  ({\smallonehalf})\,\bigl( p_x^2 + p_y^2 +p_z^2 \bigr) 
  + V_{K234}   \,, {\quad}
  V_{K234}  =  \frac{k_1}{r} + \frac{k_2}{x^2} + \frac{k_3}{y^2} + \frac{k_4}{z^2} \,,
$$
 is maximally superintegrable with a fundamental set of three  angular-momentum-related quadratic constants of motion  $(K_{J1}, K_{J2}, K_{J3})$ and a second set $(K_{4x},K_{4y},K_{4z})$ of three constants of motion of fourth order in the momenta.
\end{proposition}

\subsection{Kepler related Hamiltonian with a conformally Euclidean metric} \label{Sec42}

Now let us denote by  $\mu$ the following multiplier
\begin{equation}
 \mu  = 1/(1 -\kp/r) \,. 
\end{equation}
where $\kp$ is a real parameter. Then the new Hamiltonian given by 
 $$
 H_{K\mu} = \mu H_{K234} = \bigl(\frac{r}{r-\kp}\bigr) H_{K234} \,,{\quad} 
 \lim{}_{\kp\to 0}  H_{K\mu} =  H_{K234} \,, 
$$
takes the form 
\begin{equation}
  H_{K\mu}  =  ({\smallonehalf})\,\bigl(\frac{r}{r-\kp}\bigr)\bigl( p_x^2 + p_y^2 +p_z^2 \bigr) 
  +\frac{k_1}{r-\kp} +  \bigl(\frac{r}{r-\kp}\bigr)\Bigl( \frac{k_2}{x^2} + \frac{k_3}{y^2} + \frac{k_4}{z^2} \Bigr)\,.
\end{equation}
(In this section we denote the parameter by $\kp$ instead of $\la$ to  simplify the notation and avoid confusion with $\la_a$, and $\la_{a\mu}$, $a=x,y,z$). 

In the $\kp<0$ case the dynamics of $H_{K\mu}$  is correctly defined (the kinetic term is well defined); 
nevertheless when $\kp>0$, the Hamiltonian (and the associated  dynamics)
has a singularity at $r=\kp$; so in this case the dynamics is defined in the exterior of the sphere $r=\kp$.  
We also note that the factor $(1- \kp/r)$ shows a certain similarity with the coefficient in the Schwarzschild metric.

It is clear that $\mu$ preserves the spherical separability so the first set of the three angular momentum related functions $K_{Ja}$, $a=x,y,z$, still remain as $\kp$-independent integrals of motion for the $\kp$-dependent Hamiltonian $H_{K\mu}$; that is,  
 $$
  \bigl\{ K_{J1}\,, K_{J2}+K_{J3}\bigr\} =  0\,,{\qquad}  
  \bigl\{ K_{Ji}\,, H_{K\mu}\bigr\} =  0 \,,{\quad} i=1,2,3.
$$

Let us introduce the $\kp$-dependent functions $W_a$, $a=x,y,z$, defined as follows 
\begin{equation}
 W_x = R_x - \kp  \bigl(\frac{x}{r-\kp}\bigr)H_{K234}   \,,{\quad} 
 W_y = R_y - \kp  \bigl(\frac{y}{r-\kp}\bigr)H_{K234}  \,,{\quad} 
 W_z = R_z - \kp  \bigl(\frac{z}{r-\kp}\bigr)H_{K234}  \,, 
\end{equation}
so that they satisfy
$$
 \lim{}_{\kp\to 0}  W_x =  R_x    \,,{\quad} \lim{}_{\kp\to 0}  W_y =  R_y  \,,{\quad} 
 \lim{}_{\kp\to 0}  W_z =  R_z   \,.
$$
 Then these three functions $W_a$, $a=x,y,z$, and the above defined $\kp$-independent functions 
$$ 
 \bigl(x p_x + y p_y + z p_z\bigr)/x\,,{\quad}  
 \bigl(x p_x + y p_y + z p_z\bigr)/y\,,{\quad}   
 \bigl(x p_x + y p_y + z p_z\bigr)/z\,,
$$
are related between them pair-wise by their Poisson brackets with the Hamiltonian $H_{K\mu}$
$$
  \bigl\{ W_x\,, H_{K\mu} \bigr\} = 2 k_2 \lambda_{x\mu}\,\frac{1}{x} \Bigl(x p_x + y p_y + z p_z\Bigr) \,,{\quad}
 \Bigl\{\frac{1}{x} \bigl(x p_x + y p_y + z p_z\bigr) \,, H_{K\mu} \Bigr\} 
 = -\, \lambda_{x\mu}\,W_x \,, 
$$
$$
  \bigl\{W_y\,, H_{K\mu} \bigr\} = 2 k_3 \lambda_{y\mu}\,\frac{1}{y} \Bigl(x p_x + y p_y + z p_z\Bigr ) \,,{\quad}
 \Bigl\{\frac{1}{y} \Bigl(x p_x + y p_y + z p_z\bigr) \,, H_{K\mu} \Bigr\}  
 =  -\, \lambda_{y\mu}\,W_y \,,
$$
$$
   \bigl\{ W_z\,, H_{K\mu} \bigr\} = 2 k_4 \lambda_{z\mu}\,\frac{1}{z} \Bigl(x p_x + y p_y + z p_z\Bigr) \,,{\quad}
 \Bigl\{\frac{1}{z} \Bigl(x p_x + y p_y + z p_z\bigr) \,, H_{K\mu} \Bigr\}  
 = -\, \lambda_{z\mu}\,W_z \,,
$$
where the coefficients $\lambda_{a\mu}$, $a=x,y,z$, that are $\kp$-dependent, take the following  forms
$$
  \lambda_{x\mu} = \frac{1}{x^2}\bigl(\frac{r}{r-\kp}\bigr) \,,{\quad} 
  \lambda_{y\mu} = \frac{1}{y^2}\bigl(\frac{r}{r-\kp}\bigr)\,,{\quad}
  \lambda_{z\mu} = \frac{1}{z^2}\bigl(\frac{r}{r-\kp}\bigr) \,. 
$$
We note that if one of the three constants $k_i$ is not present in the Hamiltonian $H_{K\mu}$ then the corresponding function $W_i$ becomes invariant; for example, if $k_2=0$ then $W_x$  is  a quadratic constant of motion for $H_{K\mu}$. 

\begin{proposition}  \label{Prop7}
 Let  $M_{a\mu}$, $a=x,y,z$,  denote the following complex functions  
$$
 M_{a\mu} = W_a + {\ii} \sqrt{2 k_j}\,\frac{1}{a} \Bigl(x p_x + y p_y + z p_z\Bigr) \,,{\quad}
 a = x,y,z,{\quad} j=2,3,4 \,. 
$$
Then the Poisson bracket of everyone of these complex functions with the Hamiltonian $H_{K\mu}$  is directly related to itself and given by the following expressions 
$$
  \bigl\{ M_{x\mu} \,, H_{K\mu} \bigr\} = -\, {\ii}\sqrt{2 k_2}\, \lambda_{x\mu}\,M_{x\mu} \,,{\quad}
  \bigl\{ M_{y\mu} \,, H_{K\mu} \bigr\} = -\, {\ii}\sqrt{2 k_3}\, \lambda_{y\mu}\,M_{y\mu} \,,{\quad}
$$
$$
 \bigl\{ M_{z\mu} \,, H_{K\mu} \bigr\} = -\, {\ii}\sqrt{2 k_4}\, \lambda_{z\mu}\,M_{z\mu} \,.
$$
\end{proposition} 
 These three properties are rather similar to the previous properties in the case of the Euclidean Hamiltonian  (Proposition \ref{Prop5}) but with $\mu$-dependent functions; that is, $H_{K\mu}$ instead of $H_{K234}$,  $M_{a\mu}$ instead of $M_{a}$ and $\lambda_{a\mu}$ instead of $\lambda_{a}$, $a=2,3,4$.

  Therefore the moduli $|\, M_{a\mu}\,|$ of the three functions $M_{a\mu}$, $a=x,y,z$, satisfy 
$$
 \bigl\{ M_{a\mu} M_{a\mu}^{*} \,, H_{K\mu}\bigr\}   = 
 \Bigl(-\, {\ii} \sqrt{2 k_j}\, \lambda_{a\mu} M_{a\mu} \Bigr) M_{a\mu}^{*} 
 +   M_{a\mu}\Bigl(  {\ii}\sqrt{2 k_j}\, \lambda_{a\mu}  M_{a\mu}^{*} \Bigr)= 0 \,, 
$$
with $a=x,y,z$, and $j=2,3,4$.

Hence the three functions  $K_{4a\mu}= |\, M_{a\mu}\,|^2$,  $a=x,y,z$, given by 
$$
 K_{4x\mu}  = W_{x}^2 + \frac{2k_2}{x^2} \Bigl(x p_x + y p_y + z p_z\Bigr)^2  \,,{\quad}
 K_{4y\mu}  = W_{y}^2 + \frac{2k_3}{y^2} \Bigl(x p_x + y p_y + z p_z\Bigr)^2 \,, 
$$
\begin{equation}
 K_{4z\mu}  = W_{z}^2 + \frac{2k_4}{z^2} \Bigl(x p_x + y p_y + z p_z\Bigr)^2 \,, 
\end{equation}
are   first   integrals of motion of fourth order in the momenta.

We note that the coefficients $\lambda_{a\mu}$, $a=x,y,z$,  are not constants but functions $\lambda_{a\mu}\ne \lambda_{b\mu}$, $a\ne b$; this fact prevents the coupling of  $M_{x\mu}$ with $M_{y\mu}$ or $M_{z\mu}$. That is, we obtain the three functions $K_{4a\mu}=W_{a}^2+ \dots$ but not any function of the form $W_x W_y + \dots$

We close this section by summarizing the results in the following proposition:
\begin{proposition}  \label{Prop8}
 The $\kp$-dependent Kepler-related Hamiltonian
$$
  H_{K\mu}  = \mu H_{K234}  =  ({\smallonehalf})\,\bigl(\frac{r}{r-\kp}\bigr)\bigl( p_x^2 + p_y^2 +p_z^2 \bigr) 
  +\frac{k_1}{r-\kp} +  \bigl(\frac{r}{r-\kp}\bigr)\Bigl( \frac{k_2}{x^2} + \frac{k_3}{y^2} + \frac{k_4}{z^2} \Bigr)\,, 
$$
 is maximally superintegrable with a fundamental set of three  angular-momentum-related quadratic constants of motion  $(K_{J1}, K_{J2}, K_{J3})$ and a second set $(K_{4x\mu},K_{4y\mu},K_{4z\mu})$ of three $\kp$-dependent constants of motion of fourth order in the momenta.
\end{proposition}

In this case we present the geometric properties by making use of spherical coordinates $(r,\te.\phi)$.
The conformally flat metric $g_{ij} $  in Euclidean  coordinates  is given by 
\begin{equation}
  g_{ij} =(1 -\kp/r){\rm diagonal}[1,1,1],
\end{equation}
and then in the above mentioned spherical coordinates by 
\begin{equation}
g_{ij} =(1 -\kp/r){\rm diagonal}[1, r^2,  r^2\sin^2\theta], 
\end{equation}
and the sectional curvatures $\kappa_{r\te}$, $\kappa_{r\phi}$, and $\kappa_{\te\phi}$, with respect to the three two-dimensional  planes  $(r,\te)$, $(r,\phi)$, and  $(\te,\phi)$,  that are orthogonal to one another, take the  forms
\begin{equation}
 \kappa_{r\te} = \frac{\kp}{2(r- \kp)^3}   \,,{\qquad}
 \kappa_{r\phi} = \frac{\kp}{2(r- \kp)^3}   \,,{\qquad}
 \kappa_{\te\phi} = \frac{\kp(3\kp-4r)}{4r(r- \kp)^3}   \,, 
\end{equation}
and then the scalar curvature (Ricci curvature), that  is given by two times the sum of the three sectional curvatures, is 
\begin{equation}
 R = \frac{3 \kp^2}{2 r (r- \kp)^3} \,. 
\end{equation}
The  sectional curvatures $\kappa_{ab}$ can take positive or negative values but the Ricci curvature $R$ (that is a scalar function  representing an average of the partial curvatures) is proportional to $\kp^2$ and it is therefore always positive 
(we recall that the dynamics is defined in the region  $(r- \kp)>0$).

\section{Final comments }

As observed in the introduction the harmonic oscillator and the Kepler problem are important by themselves but also as a starting point for the study of other related but more general systems. 
In fact this has been the matter we have studied: the analysis of four 3-dimensional superintegrable systems  defined on conformally flat spaces and related with these two fundamental systems.

More precisely, we have proved the quadratic superintegrability (and we have obtained all the integrals of motion) of the following two oscillator-related Hamiltonians with a   conformally Euclidean metric
\begin{itemize}
\item Isotropic harmonic oscillator  with additional terms of the form $k_2x$, $k_3y$, and $k_4z$
$$ 
 H_\mu = \mu H_{111}  =   ({\smallonehalf})\,\Bigl( \frac{p_x^2 + p_y^2 +p_z^2}{1 - \la\,r^2} \Bigr)  
 +  \Bigl[\,  k_1\frac{x^2+y^2+z^2}{1 - \la\,r^2} + \frac{k_2\, x }{1 - \la\,r^2} +  \frac{k_3\, y}{1 - \la\,r^2} + \frac{k_4\, z }{1 - \la\,r^2}  \,\Bigr]     \,. 
$$
\item Isotropic harmonic oscillator with additional terms of the form $k_2/x^2$, $k_3/y^2$, and $k_4/z^2$
$$
 H_\mu = \mu H_{111}  =   ({\smallonehalf})\,\Bigl( \frac{p_x^2 + p_y^2 +p_z^2}{1 - \la\,r^2} \Bigr)  
 +  \Bigl[\, k_1\frac{x^2+y^2+z^2}{1 - \la\,r^2}  + \frac{k_2/x^2 }{1 - \la\,r^2} +  \frac{k_3/y^2}{1 - \la\,r^2} + \frac{k_4/z^2 }{1 - \la\,r^2}  \,\Bigr]    \,. 
$$
\end{itemize}
as well as the higher-order superintegrability (with integrals of fourth-order in the momenta) of the following 1:1:2 oscillator-related and Kepler-related Hamiltonians:
\begin{itemize}
\item Oscillator  1:1:2 with additional terms of the form  $k_2/x^2$, $k_3/y^2$, and $k_4/z^2$
$$
 H_\mu  = \mu H_{k234}  =   ({\smallonehalf})\,\Bigl( \frac{p_x^2 + p_y^2 +p_z^2}{1 - \la\, (x^2+y^2+ 4z^2)} \Bigr)  
 +  \Bigl[\,  \frac{k_1 (x^2+y^2+ 4z^2)}{1 - \la\, (x^2+y^2+ 4z^2)} + 
 \frac{k_2/x^2 + k_3/y^2+ k_4/z^2}{1 - \la\, (x^2+y^2+ 4z^2)}  \,\Bigr]     \,. 
$$
\item Kepler with additional terms of the form $k_2/x^2$, $k_3/y^2$, and $k_4/z^2$
$$
  H_{K\mu}  = \mu H_{K234}  =  ({\smallonehalf})\,\bigl(\frac{r}{r-\kp}\bigr)\bigl( p_x^2 + p_y^2 +p_z^2 \bigr) 
  +\frac{k_1}{r-\kp} +  \bigl(\frac{r}{r-\kp}\bigr)\Bigl( \frac{k_2}{x^2} + \frac{k_3}{y^2} + \frac{k_4}{z^2} \Bigr)\,.
$$
\end{itemize}

Although the initial and principal objective of this paper was the study of Hamiltonians defined on spaces with a conformally Euclidean geometry, we have arrived to an important and different question; existence of of Hamiltonian systems with higher-order integrals of motion. 
This is a very remarkable result since these integrals are not related with Hamilton-Jacobi (Schr\"odinger in the quantum case) separability and, because of this, they are very difficult to be obtained. 
In fact, most of studies devoted to this question are restricted to the two-dimensional Euclidean plane and the results for these higher order in the momenta constants are usually obtained after a long calculation.  
Here, working with 3-dimensional systems, we have obtained several quartic first integrals  by making use of a method related with the existence of complex functions which satisfy certain interesting Poisson bracket relations in such a way that the new constants do not arise from separability but from the properties of these complex functions. 

A natural question is if this complex-related method is limited to the two systems studied in sections \ref{Sec3}  and \ref{Sec4} or it can be applied to other different Hamiltonian systems. 
This is an open question that must be considered as a matter to be studied. 
We also recall that we have made use of some particular values of the multiplier $\mu$ 
(appropriate in every case to the particular expression of the potential). 
The existence of more general values of $\mu$ (more general conformally Euclidean metrics) is also a matter to be studied.

\section{Appendix. Properties of the matrix  $[K_{ij}]$}

The symmetric matrix of the $\la$-depending integrals of motion $\left[ K_{ij} \right]$,  $ \bigl\{K_{ij}\,,H_\mu\bigr\} = 0 $,  $H_{\mu} = \mu H_{111}$, obtained in the section (\ref{Sec21}) represents   a generalization of the Fradkin tensor \cite{Frad65} for the dynamics of the Hamiltonian $H_{\mu} = \mu H_{111}$. Now we present its more important properties 
(in this appendix we simplify the notation and we just write $K_{ij}$ instead of $K_{ij\la}$). 

\begin{itemize}
\item[(i)]  The trace of the matrix $[K_{ij}]$ is just the Hamiltonian
$$
 \tr[K_{ij}] = K_{xx} + K_{yy} + K_{zz} = 2 H_\mu  \,. 
$$
\item[(ii)]   The matrix  $[K_{ij}]$ satisfies the following property 
$$ 
\left[\matrix{
K_{xx}  &  K_{xy} & K_{zx} \cr
K_{xy}  & K_{yy}  & K_{yz} \cr 
K_{zx}  & K_{yz}  & K_{zz} \cr  }\right]
 \left[\matrix{J_{yz} \cr J_{zx} \cr J_{xy} }\right] = I\,\left[\matrix{x \cr y \cr z }\right] ,
$$
where we recall that $I$ is the following linear   first   integral 
$$
I  = k_2 J_{yz} + k_3 J_{zx} + k_4 J_{xy}  \,,{\qquad} \bigl\{ I \,, H_\mu\bigr\}  =0
$$
In the particular case $k_i=0$, $i=2,3,4$, the right  hand side vanishes and we obtain  the result of Fradkin.
\item[(iii)]      The following relations between the components of the matrix are true: 
$$
x^2 K_{yy} - 2 x y K_{xy} + y^2 K_{xx} =  J_{xy}^2  \,,{\qquad}
y^2 K_{zz} - 2 y z K_{yz} + z^2 K_{yy} =  J_{yz}^2  \,, 
$$
$$ z^2 K_{xx} - 2 z x K_{zx} + x^2 K_{zz} =  J_{zx}^2 ,
$$
\begin{eqnarray*}
 K_{xx}K_{yy} - K_{xy}^2 &=& 2 J_{xy}^2 (\la H_\mu+k_1) - 2 (k_3 p_x - k_2 p_y)J_{xy} - (k_3 x - k_2 y)^2 , \cr
 K_{yy}K_{zz} - K_{yz}^2 &=& 2 J_{yz}^2 (\la H_\mu+k_1) - 2 (k_4 p_y - k_3 p_z)J_{yz} - (k_4 y - k_3 z)^2.
\end{eqnarray*}
\item[(iv)]    The following three algebraic properties are true 
\begin{eqnarray*}
K_{ij} x_i x_j &=&  2 \bigl(x^2 + y^2 + z^2\bigr) H_\mu  - (J_{xy}^2 + J_{yz}^2 + J_{zx}^2),  \cr
 K_{ij} x_ip_j  &=& 2 \bigl(x p_x + y p_y + z p_z\bigr) H_\mu +  
 \Bigl[  ( k_3 x - k_2 y)J_{xy} +  (k_4 y - k_3 z) J_{yz} +  (k_2 z - k_4 x)J_{zx} \Bigr], \cr
K_{ij} p_i p_j &=&  (p_x^2 + p_y^2 + p_z^2)^2 + 2  \bigl(x p_x + y p_y + z p_z\bigr)^2 ( \la H_\mu + k_1) + 
 2 (k_2 p_x + k_3 p_y + k_4 p_z) (x p_x + y p_y + z p_z) .
\end{eqnarray*}
\end{itemize}

\section*{Acknowledgments}

J.F.C. and M.F.R. acknowledge support from research Projects No. PGC2018-098265-B-C31 (MINECO, Madrid)  and DGA-E48/20R (DGA, Zaragoza),  
M.S. acknowledges support by the research Projects No.  VA137G18 and BU229P18 (Junta de Castilla y Le\'on).

{\small
   }

\end{document}